# Hierarchical Temporal Memory Based on Spin-Neurons and Resistive Memory for Energy-Efficient Brain-Inspired Computing

Deliang Fan, Mrigank Sharad, Abhronil Sengupta, Kaushik Roy

*Abstract*—**Hierarchical temporal memory (HTM) tries to mimic the computing in cerebral-neocortex. It identifies spatial and temporal patterns in the input for making inferences. This may require large number of computationally expensive tasks like, dot-product evaluations. Nano-devices that can provide direct mapping for such primitives are of great interest. In this work we show that the computing blocks for HTM can be mapped using low-voltage, fast-switching, magneto-metallic 'spin-neurons' combined with emerging resistive cross-bar network (RCN). Results show possibility of more than 200× lower energy as compared to 45nm CMOS ASIC design.**

## I. INTRODUCTION

THE human brains are highly efficient in performing cognitive tasks which are thought to involve processing of patterns hidden in different sensory-input stimuli, followed by response-generation [1, 2]. The biological vision-system for instance, may incorporate processing of spatial/temporal patterns, the results of which may be combined with that of the auditory-system by the brain, to produce an appropriate physiological response. Several computing models have been explored in literatures [2, 3, 4] that aim to borrow from the cerebral-information-processing system, in a quest to realize 'intelligent' machines. The earliest efforts involved different mathematical models for artificial neural-networks, with varying neuron-transfer functions and connection-topologies [2]. Deep learning networks (DLN), capable of identifying patterns under large degree of spatial variations, evolved as a tool for machine learning applications of practical complexity [3]. DLNs employ a number of computing-levels, with each level processing spatially overlapping region of the inputs, thereby, leading to appreciable tolerance towards spatial-modifications of a set of 'learned' patterns [3].

Recently, temporal-processing was introduced to DLNs as an important new-feature. The resulting biomimetic computing model, called hierarchical temporal memory (HTM), offers the potential of spatial as well as temporal pattern processing, akin to the cerebral neocortex. HTM constitutes of multiple levels of processor arrays. Each processor node 'pools' spatial patterns received from the nodes in the lower-level of its 'perceptive-field' and simultaneously identifies the key temporal sequences among those spatial-patterns. The pattern-identification-process may involve computation of conventional distance metrics like, Hamming-Distance (HD), Gaussian distance (GD), or dot-product (DP) between the stored and the input patterns at each node. A practical HTM hardware may need to store and compute with hundreds of spatial/temporal patterns at every node. Implementation of such hardware, using the conventional Von-Neumann digital-architecture may incur prohibitively high energy and real-estate cost [7].

Recent years have seen growing interest in emerging nano-devices that can provide direct and energy efficient mapping of computing-primitives required for pattern matching tasks, as in HTM. The pattern matching computations, being inherently variation tolerant, can exploit the 'inexact' terminal characteristics of such nano-devices to perform non-Boolean, analog-mode operations upon inputs. More importantly, devices that can facilitate direct 'in-memory' processing, may be highly attractive for such memory intensive computing. Several device solutions have been proposed for fabricating nano-scale programmable resistive elements, generally categorized under the term 'memristor' [9-17]. Of special interest are those which are amenable to integration with state of the art CMOS technology, like memristors based on Ag-Si filaments [14-16]. Such devices can be integrated into metallic cross-bars to obtain high density resistive cross-bar networks (RCN) [9-16]. Continuous range of resistance values obtainable in these devices can facilitate the design of multilevel, non-volatile memory [9-11]. The RCN technology has led to interesting possibilities of combining memory with computation [9-13]. RCN can be highly suitable for large number of non-Boolean computing applications that involve pattern-matching [13, 19]. Such applications employ highly memory intensive computing that may require correlation of a multidimensional input data with a large number of stored

Manuscript received January 14, 2014. This work was supported in part by CSPIN, StarNet center, DARPA UPSIDE, SRC, Intel, and NSF.

The authors are with the School of Electrical and Computer Engineering, Purdue University, West Lafayette, IN, 47906, USA ( e-mail: dfan@purdue.edu; msharad@purdue.edu; asengup@purdue.edu; kaushik@purdue.edu ).

patterns or templates, in order to find the best match [19]. Use of conventional digital processing techniques for such tasks incurs prohibitively high energy and real-estate cost, due to the sheer number of computations involved. Structurally, RCN can be a much closer fit for this class of pattern matching computation. Owing to the direct use of nano-scale memory array for pattern-matching computing, it can provide very high degree of parallelism, apart from eliminating the overhead due to memory read.

Pattern matching computing of practical complexity with RCN is essentially analog in nature, as it involves evaluating the degree of correlation between inputs and the stored data. As a result, most of the designs for pattern matching hardware using RCN proposed in recent years involved analog CMOS circuits for the processing task [17, 19]. Recent experiments on analog computing with multi-level Ag-Si memristors also employed analog operational amplifiers for current-mode processing [24]. However, application of multiple analog blocks for large scale RCN may lead to power hungry designs, due to large static power consumption of such circuits. This can eclipse the potential energy benefits of RCN for non-Boolean computing. Moreover, with technology scaling, the impact of process variations upon analog circuits becomes increasingly more prominent, resulting in lower resolution for signal amplification and processing [24]. Hence, the conventional analog circuits may fail to exploit the RCN technology for energy efficient, non-Boolean computing.

The solution to this bottleneck may lie with alternate device technologies that can provide a better fit for the required non-Boolean, analog functionality, as compared to CMOS switches. Recent experiments on spin-torque devices have demonstrated high-speed switching of scaled nano-magnets with small currents [20-22]. Such magneto-metallic devices can operate at ultra-low terminal voltages and can implement current-mode summation and comparison operations at ultra-low energy cost. Such current-mode spin switches or 'neurons' can be exploited in energy efficient analog-mode computing [28, 34, 35]. In this work we present the design of RCN based energy-efficient computing blocks for HTM using such 'spin neurons'. In the proposed scheme, the spin neurons form the core of hybrid processing elements (PE) that are employed in RCN based pattern matching modules and achieve more than 200× lower computation energy as compared to conventional CMOS circuits. Application of spin neurons to RCN can therefore greatly enhance its prospect as a non-Boolean computation tool.

The rest of the paper is organized as follows. Section II describes the mathematical models for HTM. The application of RCN in non-Boolean computing and design challenges associated with a mixed-signal implementation is presented in section III. Section IV introduces the device model for spin neuron. Design of HTM computing block using spin neurons applied to RCN modules is described in section V. Section VI concludes the paper.

## II. HTM ARCHITECTURE

In this section, the basic computing architecture for HTM is described. We focus on the hardware mapping of the computing algorithm. The training process is also briefly discussed.

### A. HTM Architecture and Training

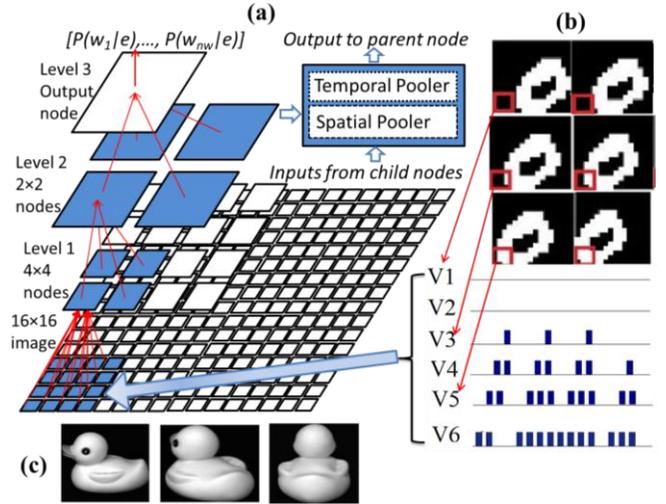

Fig. 1. (a) A three-level HTM designed to work with 16×16 pixel images (b) HTM Training Sequence generated by zigzag scan and part of the training sequence of the highlighted lower left node in level 1 (c) snap-shots of a moving duck.

The HTM computing architecture constitutes of a tree-like network of large number of processing nodes, arranged across multiple levels, having pyramidal connectivity. Each mode receives inputs from N 'child-nodes' in its 'receptive field' in the immediate lower level. The first level nodes receive inputs from an input stimulus (like, an image). Both forward as well as backward connections between the nodes of non-adjacent levels may also be used, depending upon the training algorithm and the applications [4, 5]. In this work, the specific application considered requires only feed-forward flow.

The HTM network training mainly involves the extraction of spatial and temporal patterns from the time varying input data. It proceeds from bottom to top. The parent nodes are trained only after all the child nodes in the lower levels are fully trained. All but the top-most level are trained in unsupervised mode [5, 8]. The following subsection describes the training process.

*1) Spatial Pooling*: During the training process, the HTM network is exposed to time varying inputs, such as that produced by an object moving smoothly across the network's visual field [6, 8]. Fig. 1b shows a simple training sequence generated by the moving-image of a numeric character, which may be shifting, rotating and scaling (by moving towards or away from the scanner) across the visual-field.

Training with such time-varying snap-shots of an object can help recognize it with different perspectives using a fully-trained network. A more realistic example can be given as that of a moving object, like a duck (taken from COIL-20 data-set [40]), as shown in fig.1c.

The level-1 (L1-nodes) receive M×M pixels (M=4 in this work) of the input image, which can be viewed as a 1-D spatial pattern (of length M×M). The L1-nodes detect and store the frequently recurring patterns in their receptive fields. During the training process, each spatial pattern or 'coincidence' $c_i$ is compared with the present set of patterns for similarity. It is added to the 'spatial pool' as a new pattern, if it is found to be sufficiently distinct from the existing set. The 'distinctiveness' of a new pattern, with respect to the present-set can be determined by placing a threshold on a distance-metric, like dot product (DP). This threshold can have a significant impact on the number of spatial / temporal patterns and the overall training accuracy, (as described in section-IIC). The probability of occurrence $P(c_i)$ of each spatial pattern is also stored in the form of its count of appearance during the training-process.

2) *Temporal Pooling*: Computation of the temporal patterns for a particular node involves identifying the group spatial patterns $c_i$'s that are likely to occur close in time. A 'temporal group', $g_i$, is a subset of coincidences that possibly originate from simple variations of the same 'class' of input that is smoothly moving throughout the network's receptive field [9]. Different algorithms can be used to partition the spatial patterns into a set of disjoint temporal groups $G=\{g_1,g_2,...,g_n\}$ [6,9]. In this work we employ an ad-hoc greedy algorithm for the sake of simplicity [9]. It employs a temporal activation matrix (TAC), where TAC(i,j) denotes the number of times the coincidence $c_i$ was followed by $c_j$ during the training. To start, we pick the element TAC(i,j) in the matrix with the highest value of $P(c_i) \times TAC(i,j)$. This implies selecting $c_i$ as the first element of the first temporal group. The largest non-zero value of TAC(i,j) implies that the coincidence $c_j$ has highest temporal connection with $c_i$. Hence, $c_j$ is added as the next element to current temporal group $g_i$. The next element to be added is $c_k$, where TAC(j,k) has the highest value among the elements in the row TAC(j,:). The elements already included in a temporal group $g_i$ are marked as 'assigned' and are not assigned to any other group. This recursive process terminates when the length of one temporal group exceeds the predetermined maximum group size. Thereafter, a new coincidence is selected as the beginner of a new temporal group.

3): *Computation of the matrix PCG*: The final step for training a node is the creation of PCG matrix, which essentially relates the spatial coincidence $c_i$'s of a node to its temporal groups $g_i$'s. The element $PCG(i,j)=P(c_i \mid g_j)$ denotes the conditional probability of coincidence $c_i$ given the temporal group $g_j$, or, in other words, the relative probability of occurrence of coincidence $c_i$ in the context of the group $g_j$. The elements of the PCG matrix are defined as in eq.1 [6].

$$PCG(i,j) = \begin{cases} P(c_i) & \text{if } c_i \in g_j \\ 0 & \text{otherwise} \end{cases}, \text{ for each } i=1...nc, \, j=1...ng \quad (1)$$

where, nc and ng are the maximum number of spatial patterns (coincidences) and temporal groups respectively. During the inference mode, the PCG matrix of a node is used to evaluate the probability distribution over the stored temporal groups, $g_i$'s, in that node, based on its current spatial inputs (as discussed in section-IIB). Hence, it can be termed as the 'inference matrix' of a node. The index of the temporal group with the highest probability value constitutes the output information of the node. During the training of a parent node (nodes not connected directly to the input image), all its child nodes (which are already trained), operate in the inference mode. Their outputs, (which are the indices of the winning temporal groups of the respective nodes, obtained based on current input image) form an effective spatial pattern for the parent node.

4) *Training of the output-node*: As mentioned earlier, the training steps of the output node (the node at the top of the HTM tree) is supervised. The computation of spatial pool (with elements $c_i$'s) is identical to the other levels. The inference matrix, however, is constructed through supervised learning, under a set of specified 'desirable' output classes $w_i$'s. The inference matrix of the output node can therefore be termed as the PCW matrix. The elements of the PCW matrix are updated based on the a priori knowledge of the current image class. For example, if the current input image belongs to class $w_j$, and current coincidence to the output node is identified to be $c_i$ (using DP with all $c_i$'s in the output node), the value of PCW(i,j) is incremented by 1.

## B. HTM Inference

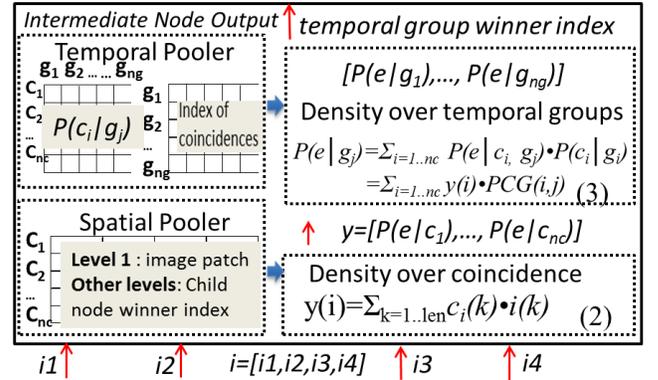

Fig. 2. HTM-node structure and the associated inference-steps

Fig. 2 shows the node structure and mathematical formulations of the inference steps used in this work [6, 9]. Inference steps for a node can be divided into the following steps:

1): *Composition of spatial input*: The spatial input to a node i=[i1,i2,...,iN] is obtained by juxtaposition of the output messages from its N child-nodes. As described earlier, for the L1 nodes, the spatial inputs are received directly from the input image (being tested). For the higher level nodes however, the spatial inputs are constituted by the winner indices of the temporal groups of their child nodes.

2): *Computation of probability densities over spatial coincidences*: The vector y shown in fig. 2 constitutes of the conditional probability distribution of the input spatial-pattern over the stored-coincidences: $y(i)=P(e|c_i)$, $i=1\cdots nc$. It encodes the spatial similarity between the input pattern (i) and the stored spatial coincidences ($c_i$'s). It can be computed as the dot product (DP) between the input (i) and the stored patterns (eq. 2 in fig. 2). A threshold operation is added after this computation, if the output value is smaller than the threshold, it will be set as zero, or it will remain the same. For the output node, a WTA circuit is needed for this step to detect the 'winner' and set the winner output to be 1, while the others to be zero.

*3): Computation of probability densities over temporal groups*: Note that, y(i) computed in step-2, denotes the probability distribution of the current spatial pattern over the pooled set of spatial coincidences ($c_i$'s). The vector PCG(:,j) on the other hand, denotes the probability of $c_i$'s, 'in context' of the particular temporal group, $g_j$. Hence, the conditional probability of the $j^{th}$ temporal group can be obtained by probability marginalization over the group coincidences shown in eq. 3 in fig. 2. We assume that $P(e|c_i,g_j)=P(e|c_i)$, since $g_j$ is irrelevant for the estimation of e density in the context of $c_i$.

*4): Computation of output message*: The output message of a node is the index of the 'winner' temporal group, which is the group with the highest value of $P(e \mid g_j)$, computed in step-3.

The inference computation of the output node is similar to the other nodes, except for the use of PCW matrix, in place of PCG matrix.

From the above discussion, we note that the core computing function for the inference mode operation of HTM is the dot product computation. At each node, this function is evaluated twice. At the first step, the operands are the analog vectors corresponding to the input spatial patterns (i(N)) and the spatial coincidences stored in the Nth node. The result, y(i), depicts the input dependent probability distribution over the pooled spatial patterns. For the second stage of computation, the input to the DP function are y(i), and, the columns of the PCG matrix, corresponding to each of the temporal groups associated with the node. The last step involves determining the index of the 'winner' temporal group, which is 'j' if the $2^{nd}$ step computing yields the highest value for DP(y(i), PCG(:,j)).

Before we move to hardware mapping of the aforementioned HTM computing scheme, we briefly discuss the choice of design specifications for HTM hardware in the following subsection.

### C. HTM Design Specification

In the previous subsections, we introduced the algorithm for training and inferring patterns using HTM, where the main computing processes involves DP-evaluation. The algorithm was applied to MNIST [39] data-set for handwritten digits recognition (fig. 3a) and COIL-20 data-set for object recognition [40]. For training, each image was scaled to 16×16 pixels and was scanned to generate a sequence of training images, incorporating a sequence of shifts, rotation and scaling of the original image. The character images were taken as binary, whereas, 4-bit resolution was chosen for the grey level COIL-20 images. As mentioned earlier, during the training process, an important parameter is the 'matching threshold' that determines the addition of a new spatial pattern to a node's memory. The relationship between the number of spatial patterns, the number of temporal groups in each node and the matching threshold are shown in figure 3b-c. These plots show that smaller threshold and hence, larger number of spatial and temporal patterns ensures higher accuracy. However, this requires increased number of DP-evaluations and hence higher

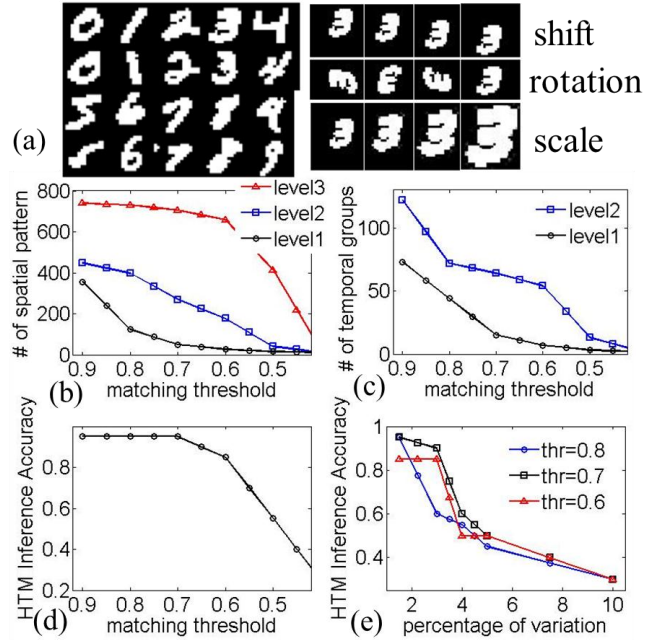

Fig. 3 (a) 20 image samples in MNIST benchmark and the shift, rotation and scale variations. (b) number of spatial patterns in each node vs. matching threshold. (c) number of temporal groups in each node vs. matching threshold. (d) HTM inference accuracy vs. matching threshold. (e) HTM inference accuracy v.s percentage-variation in the elements of spatial-temporal memory.

computation cost. In this work, the threshold was chosen close to the value for which the computation accuracy saturated to the maximum value of ~95% (threshold corresponding to 0.7). The bit-resolution required for the input and the spatial-temporal memory elements was determined by the maximum variation tolerance for which matching accuracy close to the ideal case (with non-truncated grey-scale values for memory and input) was retained (fig. 3e). During the training phase, appropriate noise models were added to the memory data and the computing function in order to account for the approximate nature of the devices-circuits characteristics used in this work.

### III. COMPUTING WITH RCN

Fig. 4 depicts a resistive cross-bar network. It constitutes of memristors (Ag-Si) with conductivity $g_{ij}$, interconnecting two sets of metal bars ($i^{th}$ horizontal bar and $j^{th}$ in-plane bar). High precision, multi-level write techniques for isolated memristors have been proposed and demonstrated in literatures that can achieve more than 8-bit write-accuracy [9, 10]. In a cross-bar array, consisting of large number of memristors, write voltage applied across two cross connected bars for programming the interconnecting memristor also results in sneak current paths through neighboring devices. This disturbs the state of unselected memristors. To overcome the sneak path problem, application of access transistors and diodes have been proposed in literature [32] that facilitate selective and disturb-free write operations. Methods for programming memristors without access transistors have also been suggested, but using such techniques, only a single device in an array can be programmed at a time [30, 31]. Such schemes can be applicable only if programming speed is not a major concern.

Memory based pattern-matching applications generally apply some form of feature reduction technique to extract and store only the essential 'patterns' or 'features' corresponding to different data samples. The extracted patterns can be represented in the form of analog vectors that can be stored along individual columns of the RCN shown in fig.4. In order to compute the correlation between an input and the stored patterns, input voltages $V_i$ (or currents $I_i$) corresponding to the input feature can be applied to the horizontal bars. Assuming the outward ends of the in-plane bars grounded, the current coming out of the $j^{th}$ in-plane bar can be visualized as the dot product of the inputs $V_i$ and the cross-bar conductance values $g_{ij}$ (fig. 4). Hence, an RCN can directly evaluate correlation between an analog input vector and a number of stored patterns. This technique can be exploited in evaluating the degree of match (DOM) between an input and the stored patterns, the best match being the pattern corresponding to the highest correlation magnitude ($\Sigma_i V_i \cdot g_{ij}$).

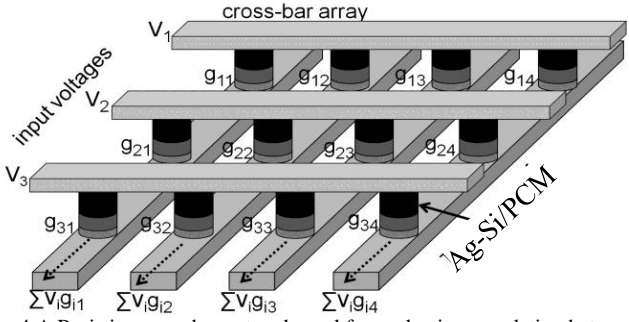

Fig. 4 A Resistive cross-bar network used for evaluating correlation between inputs and stored data.

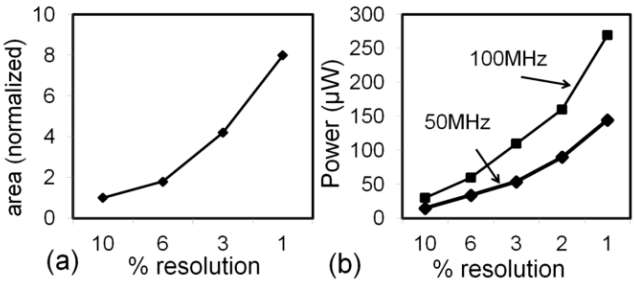

Fig. 5 Design trends using SPICE simulation: (a) higher resolution mandates larger cell area (b) for a given bias current, performance trades off with resolution and power consumption. These results were obtained using SPICE simulation of BT-WTA in [25], with σVT = 10mV for minimum sized transistors.

Variations in input source as well as memristor values were incorporated to obtain realistic values for the current outputs. For a given set of stored patterns, classification accuracy also depends upon the resolution of the detection unit used to determine the DOM for all the stored patterns. A resolution of 4% (5-bit) was chosen based on the observation that unto this value, the classification accuracy remained close to that achievable using ideal comparison (fig. 3e). Resolving ~4% difference among the current-mode dot product results requires a precision of 5-bits for the detection unit, responsible for identifying the winner pattern. A winner-take-all (WTA) circuit receives the current inputs and determines the 'winner'. Several versions of WTA circuits have been proposed in literatures, which can be classified into two broad categories: current-conveyor WTA (CC-WTA) [26], and binary tree WTA (BT-WTA) [25, 26], the latter being more suitable for large number of inputs [25, 26]. BT-WTA employs a binary tree of 2-input comparison stages which involve copying and propagating the larger of the two current inputs to the output [25].

In general, the use of such analog WTA circuits leads to large static power consumption. In fact, the power consumption of an analog WTA unit can be several times larger than the RCN itself. Moreover, performance of such current-mirror based circuits is limited by random mismatches in the constituent transistors and other non-idealities like, channel length modulation, that introduce mismatch in different current paths [24]. In order to maintain a sufficiently high resolution, larger transistor dimensions (both length as well as width) and hence, larger cell area is needed. This is evident from the recent designs [26] that although used relatively scaled CMOS technology, but inevitably employed significantly large channel lengths for such circuits. The increased parasitic capacitance coming from large channel lengths leads to lower operating frequency (fig. 5) for a given static power. Higher frequency and resolution can be achieved at the cost of increased input currents, which lead to larger power consumption [24]. Special techniques to enhance the precision of current mirrors have been proposed in literature [26], but they introduce significant overhead in terms of power consumption and area complexity. Voltage-mode processing can also be employed in RCN, however, it incurs additional overhead due to current to voltage conversion and subsequent amplifications. This leads to larger mismatch, nonlinearity and power consumption.

The above discussion suggests that the conventional mixed-signal CMOS design techniques may not be able to efficiently leverage the benefits of emerging nano-scale resistive memory technology for memory based computing. This motivates us to look towards alternate device technologies that can be more suitable for this purpose. In the next section we describe the spin based neuron model that can lead to efficient computing hardware based on RCN.

## IV. SPIN-NEURON FOR RCN

In this section, we describe the device operation of the spin based neuron model that is based on domain wall magnet (DWM) [30, 34, 45]. The circuit technique employed to interface the domain wall neuron (DWN) with CMOS units is also discussed.

In our recent works, we proposed the application of spin-torque neurons for designing ultra-low power neural networks. Applications of device structures based on lateral spin valves [28], as well as domain-wall magnets (DWM) [34, 44] were proposed.

Fig. 6a shows a three terminal spin neuron based on domain wall magnet [45]. It has a free magnetic domain d2 which forms an MTJ with a fixed magnet m1 at its top. The spin-polarity of d2 can be written parallel or anti-parallel to the two fixed spin-domain d1 and d3, depending upon the direction of current

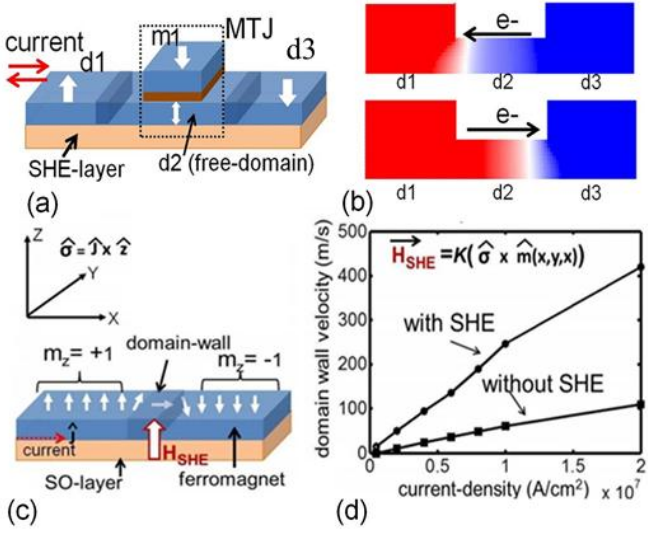

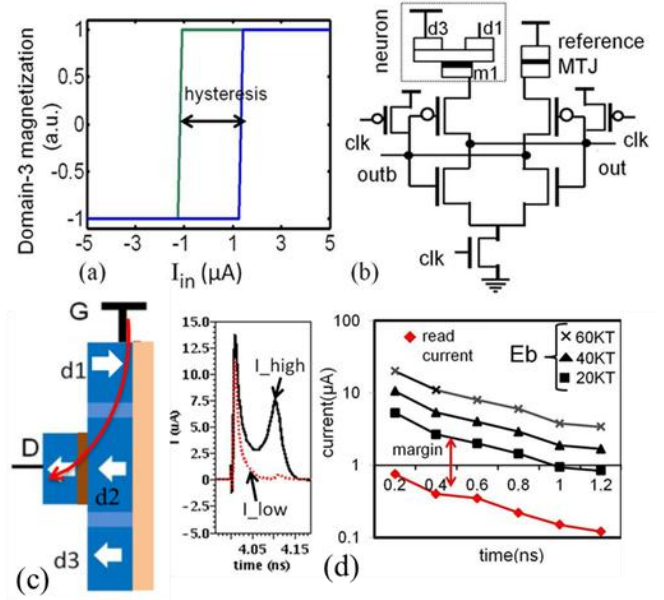

Fig. 6 (a) Three-terminal spin neuron based on domain wall magnet, (b) transient micro-magnetic simulation plots, (c) application of SHE assist for higher domain-wall (DW) speed, (d) DW-speed vs. current with and without SHE assist.

Fig. 7 (a) DWN transfer characteristics for anisotropy energy barrier, Eb=20KT, (b) dynamic CMOS latch used to detect DWN's state, (c) small transient read current for the low resistance branch can be exploited to ensure near zero read disturb, despite lowering of DW threshold. (d) The plot for read margin shows the read current for different read durations (determined by MTJ oxide thickness) and the write current required for DW motion for same durations. The read-disturb margin reduces with lowering Eb.

flow between d1 and d3. Thus, this device can detect the direction or polarity (positive if going in and negative if going out of its input domain d1) of current flow across its free domain. Hence this device can be used for current-mode thresholding operation [44]. The minimum magnitude of current flow required to flip the state of the free domain d2 depends upon the critical current density for magnetic domain wall motion across the free magnetic domain d2. Notably, domain wall (DW) velocities of ~100m/s can be reached in magnetic nano-strips with current density of ~$10^7$ A/cm² [45]. Thus a spin neuron with 60nm long fee-layer with cross-section area of $20 \times 2$ nm² may be switched with a current of less than 10µA within 1ns [45]. A non-zero current threshold for DW motion however, results in a small hysteresis in the DWN switching characteristics (fig 7a). It is desirable to reduce the threshold to get closer to the step transfer function of an ideal comparator.

A magnetic tunnel junction (MTJ) formed between a fixed polarity magnet m1 and d2 is used to read the state of d2. The effective resistance of the MTJ is smaller when m1 and d2 have the same spin-polarity and vice-versa ($R_{parallel}$~5kΩ and $R_{antiparallel}$~15kΩ). We employ a dynamic CMOS latch shown in fig. 7b to detect the MTJ state. One of its load branches is connected to the DWN MTJ whereas the other is connected to a reference MTJ whose resistance is midway between the two resistances of the DWN MTJ. The latch effectively compares the resistance between its two load branches through transient discharge currents. Note that in the detection latch, the terminal d3 of the DWN is connected to $V_{dd}$. Hence, the transient evaluation current flows from d1 to d2 as shown in fig. 7. The current required for the DW motion increases proportional to the switching current. Since the transient read current flows only for a short duration, it does not disturb the state of d2.

Robustness to read disturb can be further enhanced by the appropriate design choice of m1. Notably, the branch with effective lower resistance draws comparatively higher read current (I_high) as shown in fig. 7. By setting the polarity of m1 parallel to d3, it can be ensured that for the parallel configuration of the DWN MTJ (and hence, lower resistance) the free layer (d2) is already parallel to d3 and hence a larger transient current does not disturb d2. This technique facilitates lowering of DWN threshold to physical limits of scalability without the concern of read disturb. Apart from device scaling, the DWN threshold can also be lowered by manipulating other device parameters, like the anisotropy energy (Eb) of the magnet (fig. 7d) [44].

Recently, application of spin-orbital (SO) coupling in the form of Spin Hall Effect (SHE) has been proposed for low-current, high-speed domain-wall motion [36, 37, 38]. For Neel-type DW, SHE induced from an adjacent metal layer results in an effective magnetic-field (H$_{SHE}$) [36], that can be expressed as, H$_{SHE}$=K($\sigma \times$m). Here, m denotes the magnetization of magnetic domains. $\sigma$ is a current dependent vector defined as $\sigma = j \times z$, where, j is the current vector (which can be positive or negative depending upon direction of current flow) and z is the direction perpendicular to the magnetization plane (along easy axis). As shown in fig. 6a, $\sigma$ can be in-plan or out of plane of the figure, depending upon the direction of the current flow. K is a quantity dependent upon material parameters of the magnet and is proportional to the effective Spin Hall angle, $\theta_H$ [36]. Notably, $\theta_H$ determines the effectiveness of the Spin-Hall interaction, larger $\theta_H$ implies larger effective torque due to SHE.

For a Neel-type domain wall shown in fig. 6a, the magnetization in the region of the domain wall lies along the length of the magnetic nano-strip [36]. For this configuration, the effective $H_{SHE}$ acting on the domain wall region can be visualized to be perpendicular to the plane of the magnet. The $H_{SHE}$ assists the non-adiabatic spin-torque (which results from the current flow) acting on the domain wall region. For a $\theta_H$ of 0.2, micro-magnetic simulations showed an increase of ~5× in the domain wall velocity for a given current density, due to the $H_{SHE}$ term (fig. 6d). This effect can be used to achieve higher switching speed for a given current, or, to reduce the required switching current for a given switching time for the free domain in the spin-neuron.

In this work switching current threshold of ~2μA for 1 ns switching speed has been chosen for a neuron with SHE-assisted free domain size of $20 \times 2 \times 60 nm^3$, which corresponds to the current density of $4MA/cm^2$. This dimension of the free domain would offer an effective resistance of ~60Ω. The state of the free-domain can be sensed by injecting a small current across the high resistance magnetic tunnel junction (MTJ) formed between d2 and a fixed magnet m1.

## V. Design of HTM Computing Block Using Spin-Neuron in RCN

In the following subsections, we first describe the design of RCN based HTM computing blocks composed of two pattern matching units and their interfacing with DWNs. This is followed by circuit level description of spin-CMOS hybrid-PE based on DWN that achieves the WTA functionality at ultra-low energy cost. The implemented HTM hardware architecture is also shown in this section.

### A. Pattern Matching Network Design

Each HTM block consists of two pattern matching networks using dot product, corresponding to the computation of density over coincidences and temporal groups. The node structure and mathematical equations can be seen in fig. 2. The dot product functionality can be implemented by RCN described in section III.

As described in section-II, the dimension of each RCN based dot product computing block is (n_child×nc, nc×ng), where n_child is the number of child node, nc is the number of spatial patterns stored in current node and ng is the number of temporal groups. The input vectors to first RCN are respectively the real image pixels for level 1 node and the child node temporal group winner index for the other level nodes. The input vectors to the second RCN are the outputs of the first RCN. As shown in fig 3e, if matching threshold is chosen as 0.7, ~4% parameter variation can be tolerated. Therefore the bit-length of the PCG matrix (and of spatial pooler) was chosen to be 5. Fig. 8a depicts the DWNs with their input (d1 terminal in fig. 6a) connected to RCN outputs. A DC voltage, V, is applied to the d3 terminals of all the DWNs (access transistors and SHE layers are not shown for simplicity). Owing to the small resistance of the DWN devices, this effectively biases output ends of the RCN (connected to d1 terminals) to the same voltage. Each of the RCN digital input values is converted into analog voltages/current levels. The low voltage operation of DWN can be exploited to implement, compact and energy efficient current-mode DAC using binary weighted deep-triode

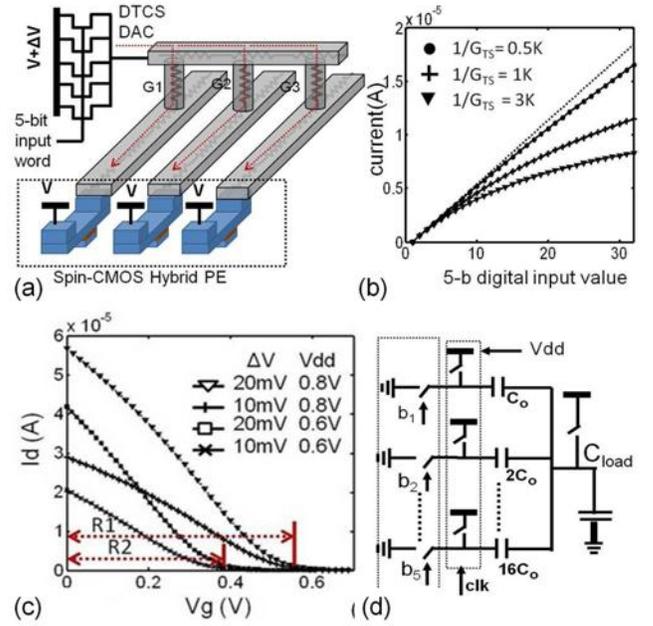

Fig. 8 (a) RCN with a single DTCS input and three receiving DWN (b) non-linear characteristics of DTCS resulting due to series combination with Gs, (c) near-linear drain-current (Id) vs. gate voltage (Vg) characteristics for DTCS, (d) analog driving scheme using compact switched capacitor DAC.

current source (DTCS) PMOS transistors, as shown in fig. 8a. DAC supply of V+ΔV is applied to the source terminals of the DTCS, where ΔV is ~30mV. Such a low value of drain to source voltage for the DTCS provide linear Id (drain-current) Vgs (gate to source voltage) characteristics that can be exploited for analog-mode driving (fig. 8c). An alternative low power DAC for the input digital data can be a compact switched MOS-capacitor DAC (fig. 8d). This analog voltage can be used to drive the DTCS transistors that supply current to the RCN for computation. Analog-mode driving can achieve lower data bus width, thereby reducing the power consumption due to dynamic switching of the data bus. The output current of each column is the dot product of the input voltages (currents) and the programmed conductance of the memristors as described in section III. The analog output currents will be converted into digital values using the proposed spin-neuron based SAR-ADC (will be described in section-VB).

Ignoring the parasitic resistance of the metal cross-bar, the drain to source voltage of the DTCS can be approximated to ΔV. The current Iin(i), supplied by the ith DTCS can thus be written as $\Delta V.GT(i)GTS/(GT(i)+GTS)$), where GT(i) is the data dependent conductance of the $i^{th}$ DTCS and GTS is the total conductance (of all the Ag-Si memristors, including the 'ON' resistance of the access transistors, if present) connected to a horizontal bar (dummy memristors are added for each horizontal input bar such that GST is equal for all horizontal bars). As a result, the current input through a memristor connecting the $i^{th}$ input bar to the $j^{th}$ output bar (in-plane) can be written as $I(i,j)=\Delta V.GT(i)GST/(GT(i)+GST)(G(i,j)/GST)$, where, $G(i,j)$ is the programmed conductance of the memristor. For accurate dot product evaluation, the current $I(i,j)$ should be proportional to the product of GT ( i.e., the DTCS conductance, proportional to the input data) and G(i,j). Hence, a low value of

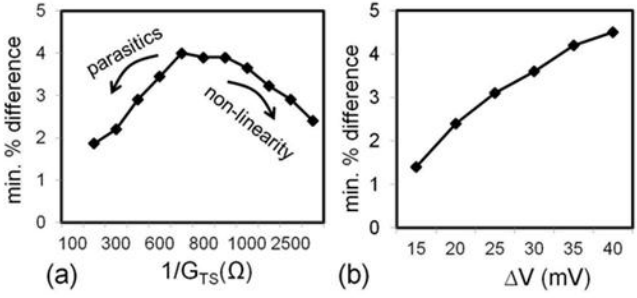

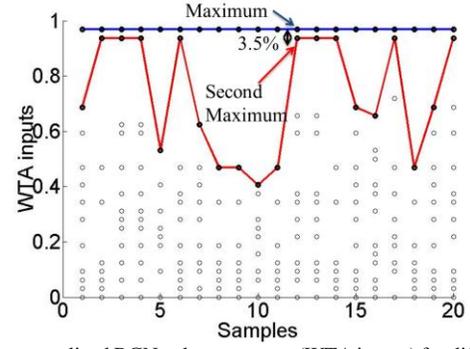

Fig. 9 (a) degradation in detection margin for a given input due to non-linearity (for low GTS) and parasitic voltage drops (for high GTS), (b) degradation in detection margin for the same input, for reducing $\Delta V$, due to parasitic voltage drops.

Fig. 10 the normalized RCN column outputs (WTA inputs) for different image samples, showing isolation between the best and second best match.

GTS (i.e. higher resistance values of the memristors) introduces non-linearity in the DTCS-DAC characteristics (fig. 8b). This leads to reduction in the detection margins (difference between the best and the second best match) for the current-mode dot product outputs for different input vectors (fig. 9a). As a result, the overall matching accuracy of the network reduces for a given WTA resolution. Ideally, choosing the lowest possible range of values for the memristor resistances (say 200Ω-6.4K Ω, no access transistor being used) would largely overcome the non-linearity (fig. 9). However, for higher G(i,j), voltage drop in the metal lines due to parasitic resistances result in corruption of the current signals, once again, leading to degradation in the detection-margin. Hence, the optimal range for the conductance values was found based on the maximum achievable read margin, as shown in fig. 9a. Note that, in case access transistors are employed for improved writablity, the minimum conductance is determined by the ON-resistance of the transistors (which is ~1K Ω for a minimum sized 45nm device). The Ag-Si memristors can be programmed to low resistance value of ~100 Ω. The design parameters like the matching threshold, data bit-width etc., discussed earlier, were therefore determined based on the simulation of RCN model, in order to ensure resolvable detection margin.

## B. Winner Take All Design

As described in section-II, the output of each node in HTM is the winner index of the temporal group for the non-output node or the winner index of the class for the output node. A winner take all (WTA) circuit should be attached to the RCN based pattern matching network. The DWN device essentially acts as a low-voltage, high-speed, high-resolution current-mode comparator and hence can be exploited in digitizing analog current levels at ultra-low energy cost [28]. The proposed WTA scheme, algorithmically depicted in fig. 11, exploits this fact and clubs a digitization step with a parallel 'winner-tracking' operation.

Fig.10 shows the normalized RCN output (WTA inputs) for the HTM level 2 nodes, for 20 different image samples. It shows the worst-case separation between the best and the second-best matches to be ~4%, at the moment of comparison, which indicates at least a 5 bit resolution WTA circuit is needed to detect the best match.

The first half of the flowchart (fig. 11) can be identified as the standard algorithm for successive approximation register

(SAR) ADC [28]. The data conversion algorithm employed in an SAR-ADC can be explained as follows. To begin the conversion, the approximation register (that stores the digitization result) is initialized to the mid-scale (i.e., all but the most significant bit is set to 0). At every cycle a digital to analog converter (DAC) produces an analog level corresponding to the digital value stored in the SAR and a comparator compares it with the analog input using an analog comparator. If the comparator output is high, the current bit remains high, else it is turned low and the next lower bit is turned high. The process is repeated for all the bits. At the end of conversion, the SAR stores the digitized value corresponding to the analog input.

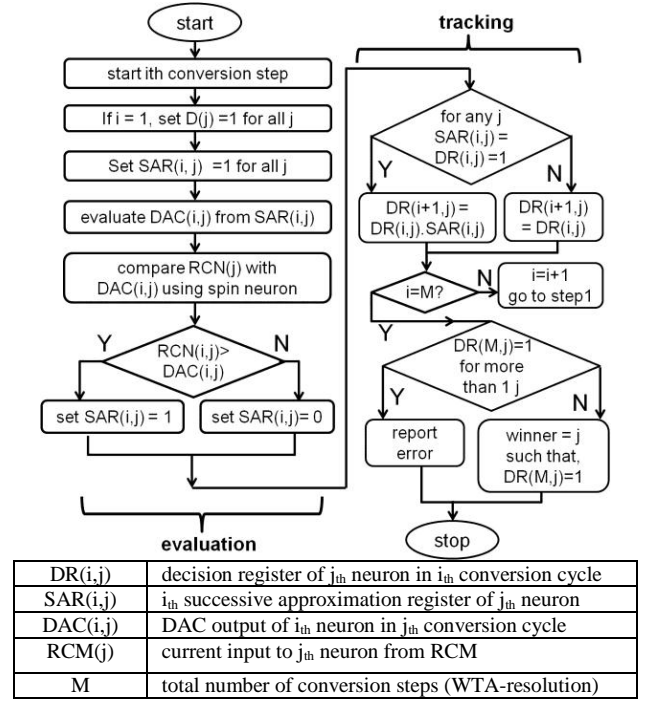

| DR(i,j) | decision register of $j_{th}$ neuron in $i_{th}$ conversion cycle |
|---|---|
| SAR(i,j) | $i_{th}$ successive approximation register of $j_{th}$ neuron |
| DAC(i,j) | DAC output of $i_{th}$ neuron in $j_{th}$ conversion cycle |
| RCN(j) | current input to $j_{th}$ neuron from RCM |
| M | total number of conversion steps (WTA-resolution) |

Fig. 11 WTA algorithm used in this work

The circuit realization of this operation using DWN's is shown in fig. 12. Output currents of the RCN columns (columns corresponding to temporal groups in level 1 and 2 nodes) are received by individual DWN input nodes that are effectively clamped at a DC supply V, as described earlier. Each DWN has an associated DTCS-DAC, which is driven by the corresponding successive approximation register. The drain terminals of the DTCS transistors are at a DC voltage V−ΔV. In each conversion cycle, the DWN device essentially compares the RCN output and the DAC output (and hence, acts as the comparator of the SAR block). The comparison result is detected by the latch described in fig. 7b, and the result is used to modify the SAR logic using the scheme described above (though pass-gate based multiplexers P, driven by a global controller). Note that, in the overall scheme, the component of RCN output current sunk by the DTCS in the ADC's flow through across a DC level of 2ΔV.

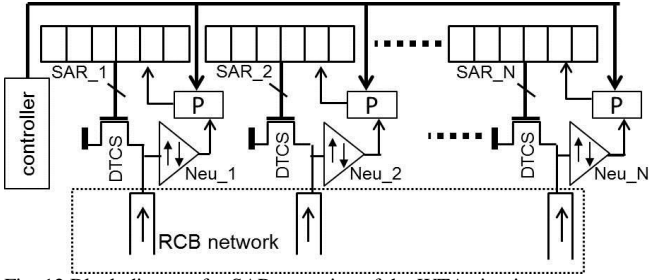

Fig. 12 Block diagram for SAR operation of the WTA circuit

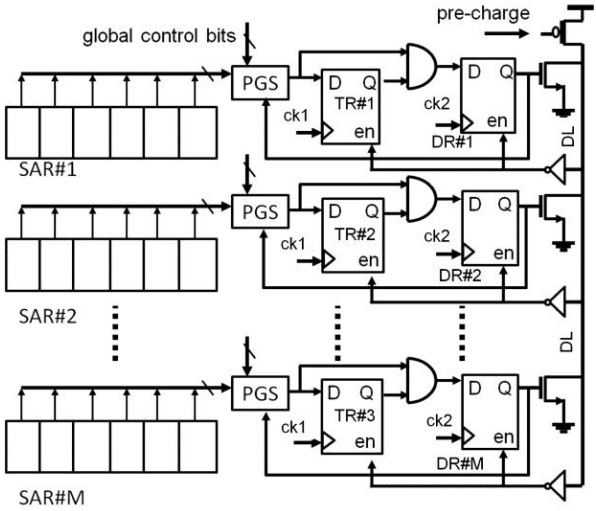

Fig.13 Circuit operation for the tracking part of the WTA algorithm

The second half of the WTA algorithm operates in parallel with the first (i.e., the ADC operation). It can be explained with the help of the corresponding circuit diagram shown in fig. 13. Results of the first ADC conversion step obtained from the SAR are directly transferred to the tracking registers (TR) shown in the figure through the pass-gate multiplexing switch (PGS). Thus, at this stage, all the TR's with a high output correspond to the ADC results with MSB = '1'. Let us now, consider the second cycle operation. The detection line (DL) is first pre-charged to Vdd and the set of discharge registers (DR) driving it are cleared to low output. Next, if for at least one of the SAR's with high MSB, the second MSB also evaluates to '1', the corresponding DR is driven high by the associated AND gate. Thus, DL is discharged to ground and the write of all the TR's is enabled. All the TR's for which both, first and second MSB's evaluated to '1', stay high, but the rest are set to low. In simple terms, if at least one of the SAR's (5-bit) evaluated to '11000' in the second conversion cycle, the DL is discharged and all the TR's with SAR value '11000' stay high, while those with SAR value '10000' are set to low. In case all SAR's evaluated to '10000' in the second cycle, no change is made to the TR values. Thus, at the end of conversion cycle, if only one of the TR's remains high, it is identified as the winner and the corresponding SAR value is effectively the degree of match (DOM). In case a random image is input to the hardware, the proposed scheme will still identify the 'winning' pattern. But if the DOM is lower than a predetermined threshold, the winner is discarded, implying that the input image does not belong to the stored data set.

The winner-tracking circuitry described above is fully digital. Moreover, owing to the global digital control, it is easily scalable with number of input as well as required bit precision.

## C. HTM Hardware Mapping Using Spin-RCN Based Pattern Matching Network Architecture

We introduced the design of RCN based pattern matching network, spin-neuron based SAR-ADC and WTA in the previous subsections. The architecture of the HTM system and each node design is shown in fig. 14. The level 1 node take the corresponding image patch as the inputs, the first RCN computes the density over spatial patterns, the SAR-ADC converts the current outputs into digital value and sends to the next RCN that computes the density over the temporal groups. The Spin-WTA circuit detects the winner and sends the winner index to its parent node. In the MNIST [39] digit recognition application, the output node will identify which digit it is for the current input image based on the WTA output.

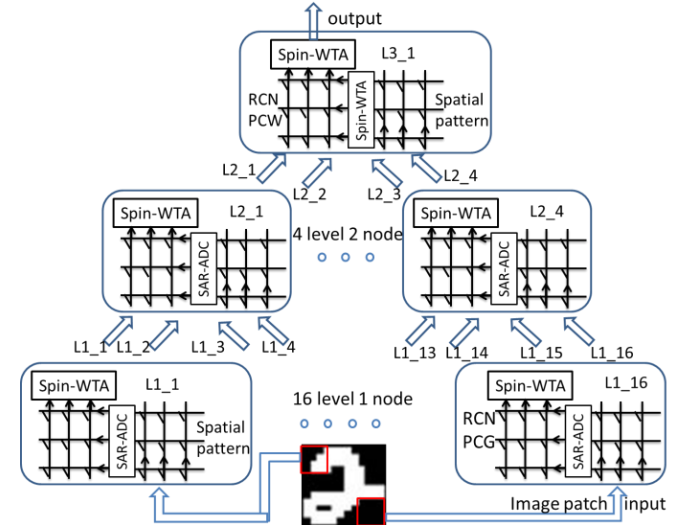

Fig. 14 HTM hardware mapping using spin-RCN based pattern matching network architecture

## D. Performance and Prospects

In order to compare the performance of the proposed HTM hardware design with state of the art CMOS design, we simulated the HTM node using digital CMOS adders and multipliers in IBM 45nm technology. The overall energy consumption in the proposed HTM hardware mapping is drastically reduced as compared to a CMOS realization, due to two main reasons; firstly, the energy consumption in the RCN itself is significantly lowered due to low-voltage operation, and secondly, the fully digital WTA scheme avoids any additional static power consumption. Note that the proposed WTA

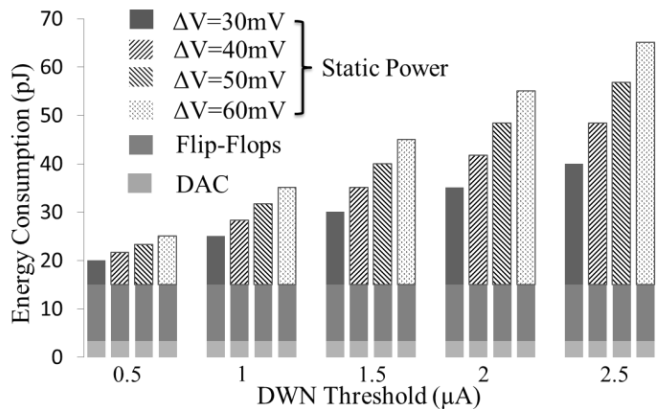

Fig. 15 For a single HTM node (level 2): Energy consumption of the proposed design with its static and dynamic components for different values of DWN threshold and different delta-V

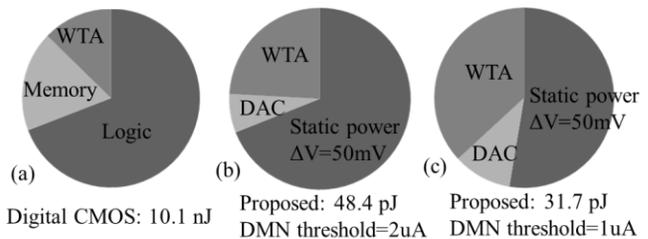

Fig. 16 Distribution of energy dissipation for a single HTM node design (level 2 node) (a) fully digital CMOS design, (b) Spin-RCN based design with 2 µA DWN threshold, (c) Spin-RCN based design with 1 µA DWN threshold ('WTA' in the pie chart includes both the ADC and WTA circuit )

scheme implemented in Mixed-Signal (MS)-CMOS would result in large power consumption, resulting from conventional ADC's. The low-voltage current-mode switching characteristic of DWN however, provides a compact and ultra-low power digitization technique [28].

As shown in fig 3, for appreciable matching accuracy, the average number of spatial ($c_i$'s) as well as temporal ($g_i$'s) groups in the HTM nodes can be more than hundred (for the given application and tree structure). As an example, for most second level nodes, the size of the PCG matrix was found to be ~270×64. This would imply DP evaluation between 64 pairs of analog vectors, each of length 270. Here, 270 denotes the length of y(i) and that of the PCG columns (PCG(:,j), each corresponding to a particular temporal group $g_j$). The bit-length of the PCG matrix (and of spatial pooler) was chosen to be 5 (based on the analysis presented in section-II). This calls for more than ~10kB of memory read per cycle of a node's computation. (If a fully parallel design is chosen for the node, it would require, storing of the same amount of data in dedicated registers). CACTI simulations [41] predict more than ~1nJ of energy dissipation, even if zero leakage digital spin-memory is used. The digital data corresponding to the PCG elements needs to be converted into analog voltage (current) levels, before it is subtracted from the analog mode results for y(i). This energy was estimated to be ~70pJ for approximate switch capacitor based DACs used in [34].

Let us now consider the energy dissipation of the proposed computing core of HTM. The simulation shows that the energy consumption of the SAR-ADC and WTA units dominates the total energy. Since there is negligible static power consumption in the WTA operation, the energy dissipation for the spin-neuron is the dominant part. The energy dissipation for the spin-neuron has two components. The first is switching energy due to the static current flow between the input voltages and the neuron. This component equals to the product of the total input current flowing across the RCN output columns, the input voltage levels, and the neuron switching time. For an average of 50µA of current flow across input voltage levels of 50mV for 1 ns switching time, this component evaluates to 2.5fJ. The noise considerations in the state of the art on-chip supply distribution schemes may limit the minimum input voltage levels that can be used. Even for 100mV of input levels, the first energy component is limited to 5 fJ. The second component of energy dissipation in the spin-neuron can be ascribed to the MTJ-based read operation. A read current of 0.3 µA (~10% of neuron switching threshold) was found to be sufficient for 1 ns read-speed. For a supply voltage of 0.8V, this would evaluate to 0.24 fJ. Thus, the total energy-dissipation in a spin-neuron for 1 ns switching speed can be around 3fJ.

Fig. 15 shows the energy consumption of a single HTM level 2 node design, where the static power consumption in the DWN-based design can be significantly reduced by lowering the DWN switching threshold further. However, the dynamic power remains constant and starts to dominate for reduced DWN thresholds. In this work, we choose the threshold current to be 2 µA. Lower value of $\Delta V$ would imply higher energy savings. We have assumed that regulated precision DC levels with ~1mV accuracy are available [43]. The minimum usable $\Delta V$ is limited by the precision regulation of DC supply achievable. As, for the given application, the required bit-precision for the spatial-temporal memory was found to be 5bit. Hence, even a 1mV noise would mandate a minimum $\Delta V$ of ~30mV. We choose $\Delta V$ as 50mV in this work to obtain better variation tolerance. Under the current DWN threshold and $\Delta V$ configuration, fig. 16 shows for a single HTM level 2 node design, the energy dissipation of the proposed design is around 48pJ, implying an energy benefit of more than 200× over a digital CMOS design. As mentioned earlier, IBM 45nm technology was used to evaluate the CMOS design energy consumption.

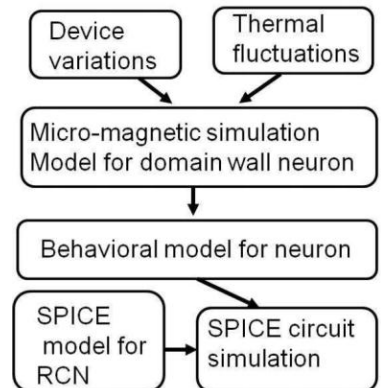

Fig. 17 simulation framework used in this work

A self-explanatory pictorial depiction of the simulation framework used in this work is given in fig. 17. We used micro-magnetic simulation model for DWN that was calibrated with experimental data on DWMs [30]. Effect of thermal fluctuation and device heating were also considered for characterizing the device. Behavioral model based on statistical characteristics of the device were used in SPICE simulation to assess the system level functionality. Some important design parameters used are listed in table I.

TABLE I
DESIGN PARAMETERS

| WTA resolution | 5 bit | Magnet material | NiFe |
|---|---|---|---|
| Input data rate | 100 MHz | Free layer size | $20 \times 2 \times 60$ nm$^3$ |
| Crossbar parasitic | 1$\Omega$/$\mu$m 0.4fF/$\mu$m | Ms | 800emu/cm$^3$ |
| Crossbar material | Cu | Ku2V | 20KT |
| Memristor material | Ag-Si | Ic | 2$\mu$A |
| Resistance range | 1K$\Omega$ to 32K$\Omega$ | Tswitch | 1ns |

## VI. CONCLUSION

The low-voltage, fast-switching, magneto-metallic 'spin-neurons' combined with RCN are explored in the dot product based pattern matching, which is the core computing block in the design of HTM hardware. Such a direct mapping of the core-computing primitive of the cortical computing system can be very attractive for large-scale and energy-efficient design. The resulting design can achieve more than 200× lower energy cost as compared to a dedicated digital hardware.

## ACKNOWLEDGMENT

The research reported here was funded in part by CSPIN StarNet center, DARPA UPSIDE, SRC, Intel, and NSF.

## REFERENCES

[1] D.C. Van Essen et. al, "Information Processing in the Primate Visual System: An Integrated Systems Perspective", *Science*, vol. 255, 1992
[2] R. Lippmann et. al, "An Introduction to Computing with Neural Nets", *IEEE ASSP Magazine*, April, 1987
[3] P. Simard et. al, "Best Practices for Convolutional Neural Networks Applied to Visual Document Analysis", ICDAR, 2003
[4] D. George and J. Hawkins, "A Hierarchical Bayesian Model of Invariant Pattern Recognition in the Visual Cortex", *IJCNN*, 2005.
[5] D. George and J. Hawkins, "Towards a Mathematical Theory of Cortical Micro-circuits", PLoS Computational Biology, 5(10), 2009.
[6] D. Maltoni et. al., "Pattern Recognition by Hierarchical Temporal Memory", University of Bologna technical report, April 2011
[7] W. Melis et. al, "Evaluation of the Hierarchical Temporal Memory as Soft Computing Platform and Its VLSI Architecture", ISMVL, 2009
[8] D. George et. al., "The HTM Learning Algorithms", Numenta tech. report, March 1, 2007
[9] S. Shin et. al, "Memristor-Based Fine Resolution Programmable Resistance and Its Applications", ISCAS, 2009.
[10] R. Berdan et.al., "High precision analogue memristor state tuning" *Electronics Letters*, 2012
[11] Feng Miao et al., "Continuous Electrical Tuning of the Chemical Composition of TaOx-Based Memristors" *ACS Nano*, 2012
[12] K. Likharev et. al.,"Biologically Inspired Computing in CMOL CrossNets", 2009
[13] J. Turelsi. al. , " Neuromorphic architectures for nanoelectronic circuits", *Int. J. Circ. Theor. Appl.* 2004

[14] S. H. Jo et.al., "High-Density Cross-bar Arrays Based on a Si Memristive System", *Nano Letters*, 2009
[15] S. H. Jo et.al, "CMOS Compatible Nanoscale Nonvolatile Resistance Switching Memory", ASC, 2008
[16] L. Gao et. al., "Analog-Input Analog-Weight Dot-Product Operation with Ag/a-Si/Pt Memristive Devices", VLSISOC, 2012
[17] B. Mouttet, " Proposal for Memristors in Signal Processing", *NanoNet* ,2009
[18] D. Morris et. al., "mLogic: Ultra-Low Voltage Non- Volatile Logic Circuits Using STT- MTJ Devices" DAC, 2012
[19] M. Hu et. al., "Hardware Realization of BSB Recall Function Using Memristor Cross-bar Arrays" , DAC 2012
[20] C. K. Lim, "Domain wall displacement induced by subnanosecond pulsed current", *App. Phy. Lett.*, 2004
[21] J. Vogel et. al., "Direct Observation of Massless Domain Wall Dynamics in Nanostripes with Perpendicular Magnetic Anisotropy", arXiv:1206.4967v1, 2012
[22] Ngo et. al., " Direct Observation of Domain Wall Motion Induced by Low-Current Density in TbFeCo Wires", *Applied Physics Express*,2011
[23] S. Fukami et al., "Low-current perpendicular domain wall motion cell for scalable high-speed MRAM", VLSI Tech. Symp, 2009
[24] P. R. Kinget, "Device Mismatch and Tradeoffs in the Design of Analog Circuits" *JSSC*, 2005
[25] Andreas et. al., " A CMOS Analog Winner-Take-All Network for Large-Scale Applications" , *IEEE TCAS*, 1998
[26] D.Lugosz et. al., " Low power current-mode binary-tree asynchronous Min/Max circuit", *Microelectronics Journal*, 2009.
[27] Sani R. Nassif, "Process Variability at the 65nm node and Beyond" CICC, 2008
[28] M. Sharad, et. Al, "Spin Neurons for ultra low power computational hardware", DRC , 2012
[29] K. Kim et. al, "A Functional Hybrid Memristor Cross-bar- Array/CMOS System forData Storage and Neuromorphic Applications", *Nano Letters*,2011
[30] C. Augustine et al., "A Self-Consistent Simulation Framework for Spin-Torque Induced Domain Wall Propagation", IEDM , 2011
[31] C. Jung et al., "Two-Step Write Scheme for Reducing Sneak-Path Leakage in Complementary Memristor Array", *IEEE Trans. on Nanotech.*, 2012
[32] H. Manem et al., "A Read-Monitored Write Circuit for 1T1M Multi-Level Memristor Memories", ISCAS, 2012.
[33] R. O. Duda, et al "Pattern Classification", JohnWiley & Sons: 2000.
[34] M.Sharad et. al, "Ultra Low Power Associative Computing Using Spin Neurons and Resistive Cross-bar Memory", DAC,2013
[35] M.Sharad et. al, "Spin-neuron: A Possible Path to Energy-Efficient Neuromorphic Computers", *Journal of Applied Physics*, 2013
[36] A. V. Khvalkovskiy, "Matching domain-wall configuration and spin-orbit torques for efficient domain-wall motion", *Physical Review*, 2013
[37] I. M. Miron, et al, "Fast current-induced domain-wall motion controlled by the Rashba effect ", *Nature Materials*, 2011
[38] KW. Kim, et al, "Magnetization dynamics induced by in-plane currents in ultrathin magnetic nanostructures with Rashba spin-orbit coupling", *Physical Review*, 2012
[39] http://yann.lecun.com/exdb/mnist/
[40] http://www.cs.columbia.edu/CAVE/software/softlib/coil-20.php
[41] http://www.hpl.hp.com/research/cacti/
[42] Z. Zeng et al., "Ultralow-Current-Density and Bias-Field-Free Spin-Transfer Nano-Oscillator", Nature Scientific Reports, 2013
[43] RC. Kuo et. al., "A High Precision Low Dropout Regulator with Nested Feedback Loops", APCCAS, 2010
[44] M. Sharad et al., "Boolean and Non-Boolean Computing using Spin Neurons" IEDM, 2012
[45] J. A. Currivan, et. al "Low energy magnetic domain wall logic in short narrow ferromagnetic wires", *IEEE Magnetics Lett.* 3, 2012.

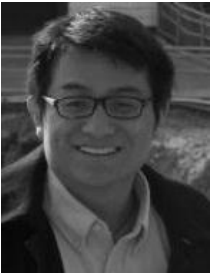

**Deliang Fan** received his B.S. degree in Electronic Information Engineering from Zhejiang University, China, in 2010 and M.S. degree in Electrical and Computer Engineering from Purdue University, IN, USA, in 2012.Currently he is a graduate research assistant of Professor Kaushik Roy and pursuing Ph.D. degree in Electrical and Computer Engineering at Purdue University.

His primary research interest lies in cross-layer (algorithm/architecture/circuit) co-design for low-power Boolean, non-Boolean and neuromorphic computation using emerging technologies like spin transfer torque devices. His past research interests include cross-layer digital system optimization and imperfection-resilient scalable digital signal processing algorithms and architectures using significance driven computation.

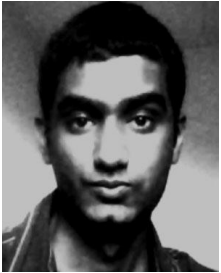

**Mrigank Sharad** received the Bachelor's and Master's degree in electronics and electrical communication engineering from IIT Kharagpur, India, in 2010, where he specialized in Microelectronics and VLSI Design. Currently he is working toward the Ph.D. degree in electrical and computer engineering at Purdue University.

His primary research interests include low-power digital/mixed-signal circuit design. His current research is focused on device-circuit co-design for low power logic and memory, with emphasis on exploration of post-CMOS technologies like, spin-devices. He has pioneered the concept of spin-CMOS hybrid design for ultra-low power neuromorphic computation architectures. He has also worked on application of spin-torque devices in approximate computing hardware and memory design.

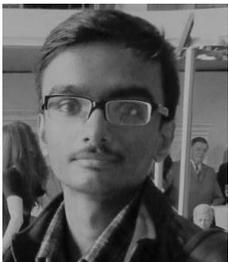

**Abhronil Sengupta** received the Bachelor's degree in electronics and telecommunication engineering from Jadavpur University, India, in 2013. Presently he is pursuing Doctoral studies as Birck Fellow and Graduate Research Assistant in the School of Electrical and Computer Engineering at Purdue University from Fall 2013.

His primary research interests lies in low-power neuromorphic computing using spin-devices.

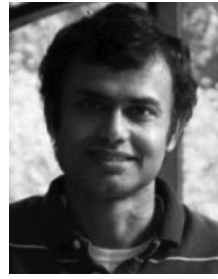

**Kaushik Roy** (F'02) received the B.Tech. degree in electronics and electrical communications engineering from Indian Institute of Technology, Kharagpur, India, and the Ph.D. degree from the University of Illinois at Urbana-Champaign, Champaign, in 1990.

He was with the Semiconductor Process and Design Center, Texas Instruments, Dallas, where he worked on field-programmable gate array architecture development and low-power circuit design. In 1993, he joined the Faculty of Electrical and Computer Engineering, Purdue University,West Lafayette, IN, where he is currently a Professor and holds the Roscoe H. George Chair of Electrical & Computer Engineering. In 2002, he was a Research Visionary Board Member of Motorola Labs. He is the author of more than 500 papers in refereed journals and conferences and a coauthor of two books on low-power complementary metaloxidesemiconductor very large scale integration (VLSI) design (John Wiley andMcGrawHill).He is the holder of 15 patents.He has graduated 50 Ph.D. students. His research interests include spintronics, VLSI design/computer-aided design for nanoscale silicon and nonsilicon technology, low-power electronics for portable computing and wireless communications, VLSI testing and verification, and reconfigurable computing.

Dr. Roy was the recipient of the National Science Foundation Career Development Award in 1995, the International BusinessMachine Faculty Partnership Award, theAT&T/Lucent FoundationAward, the 2005 Semiconductor Research Corporation (SRC) Technical Excellence Award, the SRC Inventors Award, the Purdue College of Engineering Research Excellence Award, the Humboldt Research Award in 2010 and the best paper awards at the 1997 International Test Conference, the IEEE 2000 International Symposium on Quality of Integrated Circuit Design, the 2003 IEEE Latin American Test Workshop, the 2003 IEEENANO, the 2004 IEEE International Conference on Computer Design, the 2006 IEEE/Association for Computer Machinery International Symposium on Low Power Electronics & Design, the 2005 IEEE Circuits and System Society Outstanding Young Author Award (Chris Kim), and the 2006 IEEE TRANSACTIONS ON VLSI SYSTEMS best paper award. He has been in the editorial board of IEEE DESIGN AND TEST, IEEE TRANSACTIONS ON CIRCUITS AND SYSTEMS, and IEEE TRANSACTIONS ON VLSI SYSTEMS. He was a Guest Editor for a special Issue on low-Power VLSI in the IEEE Design and Test in 1994, the IEEE TRANSACTIONS ON VLSI SYSTEMS in June 2000, and the IEEE ProceedingsComputers and Digital Techniques in July 2002.